\documentclass[]{aa}
\usepackage{graphicx}
\usepackage{amssymb,amsmath}
\usepackage{threeparttable}
\usepackage{txfonts}
\usepackage{float}
\usepackage{subfigure}
\usepackage{tabularx}
\usepackage{epsfig}

\begin{document}

\title{AGB populations in post-starburst galaxies}

\author{Jorge Melnick\inst{1} \and Roberto De Propris\inst{2} }
\institute{European Southern Observatory, Av. Alonso de Cordova 3107, Santiago, Chile
    \and
Finnish Centre for Astronomy with ESO, University of Turku, V{\"a}is{\"a}l{\"a}ntie 20,    
   Piikki{\"o}, 21500, Finland}

\offprints{Jorge Melnick \email{jmelnick@eso.org}}
\date{}

\authorrunning{Melnick  \& De Propris}

\titlerunning{Models of K+A galaxies}

\label{firstpage}

\abstract{In a previous paper we compared the SEDs of a sample of 808 K+A galaxies from the FUV to the MIR to the predictions of the 
spectrum synthesis models explicitly using AGB components. Here we use the new AGB-light models from C. Maraston  (including less fuel for the later stages of stellar evolution and
improved calibrations) to address the discrepancies between our observations and the AGB-heavy models used in our previous paper, which over-predict the infrared fluxes of post-starburst galaxies by an order of magnitude. The new models 
yield a much better fit to the data, especially in the near-IR, compared to previous 
realizations where AGB stars caused a large excess in the H and 
K bands. We {  also compare the predictions of the M2013 models to those with BC03 and find that both reproduce the observations equally well. }
We still find a significant discrepancy with {  both sets of models} in the Y and J bands, which however is
probably due to the spectral features of AGB stars. We also find that {  both the M2013 and the BC03 models} still over-predict the observed fluxes in the UV
bands, even invoking extinction laws that are stronger in these bands. While there may be some simple explanations
for this discrepancy, we find that further progress requires new observations and better modelling. 
Excess mid-infrared emission longward of 5$\mu$m is well modelled
by a $T_{dust}=300^oK$ Black-Body, which may arise from dust emission from the circumstellar envelopes
of Oxygen rich M stars (expected for a metal-rich population of AGB stars).}

 
\maketitle
\section{Introduction}
\label{intro}

Post-starburst galaxies (sometimes also called E+A or K+A galaxies) have undergone a
significant  episode of star formation (\citealt{Kaviraj2007,
Choi2009}; \citealt{Melnick2013}, hereafter paper I), but now 
show no evidence that they are currently forming stars \citep{Shin2011,Nielsen2012}, 
leaving behind an intermediate-age ($\sim0.5$ Gyr) stellar population (see 
\citealt{Quintero2004} for a review of their properties and scenarios for their formation). 
They are the best local examples of quenched galaxies and allow us to understand how 
formerly star-forming galaxies cross the 'green valley' to join the red sequence and explain
the persistent bimodality in galaxy properties across a wide range of environments \citep{Wetzel2012}.

A particular aspect we wish to emphasise here is the possibility of using these objects as calibrators of  the effects of the Asymptotic Giant Branch (AGB) phase of stellar evolution on the integrated  spectral energy distribution of galaxies \citep{Maraston1998,Lancon1999}. Models that include  the AGB phase for intermediate-mass stars yield redder colours (and lower stellar masses) for 'young' galaxies at high redshift \citep{Maraston2005,Tonini2009,Tonini2010} and may help to  solve the apparent discrepancy between  observations suggesting the presence of old and massive galaxies at $z \sim 2$ and beyond \citep{Yan2004,Cirasuolo2010,Henriques2011}, and simulations where such objects are not easily formed at high look-back times.  As post starburst galaxies contain a <0.5 Gyr old population comprising a large fraction of their current light emission, these objects may be suitable counterparts to the putative young red galaxies at high redshifts, that are also dominated by intermediate-age stars,

In Paper~I we explored these issues by using the SDSS DR7 \citep{Abazajian2009} spectra for 808 K+A
galaxies to model their stellar populations using {\sc Starlight}  \citep{Cid2005}, and  compared the results to 
spectral energy distributions from archival data spanning from 0.15 to 22 $\mu$m. We did not detect 
the luminosity boosting in the infrared expected from AGB stars in the models used by \cite{Tonini2009,
Tonini2010}. This is in agreement with the observations of post-starburst galaxies at $0.7 < z < 2$ by
\cite{Kriek2010} and the lack of strong AGB features in the infrared spectra of $z\sim0.2$ K+A galaxies 
in \cite{Zibetti2013}, although we shall argue below that their spectra miss the most important putative 
AGB features.  We also found some puzzling features, especially the lack of any correlation between model 
predictions and the metallicities derived from the spectra, suggesting that at least some of the discrepancies 
may be attributable to issues with the input physics or the reference spectra of AGB stars.

On the other hand, \cite{Miner2011} observe the predicted AGB features in the spectrum of the local
K+A NGC 5102. \cite{Eminian2008} find red infrared colors for low redshift star-forming galaxies. \cite{
Macdonald2010} show that a Thermally Pulsing (TP) AGB contribution is necessary to reproduce the  
broadband colors of spiral galaxy bulges. \cite{Melbourne2012} observe that AGB stars contribute a 
significant fraction of the infrared flux in resolved color-magnitude diagrams of nearby starburst and 
post-starburst galaxies.  Reviewing the available evidence \cite{Conroy2010} conclude that models where 
the contribution from the AGB phase is less prominent (i.e., less fuel is available for these stars) are more
likely to provide a better match to the observations than the original \cite{Maraston2005} models 
(although see \citealt{Raichoor2011}).

In our data we also observed an excess of near and far ultraviolet light over the models, as well as emission at 
$\lambda > 5$ $\mu$m significantly above the expected behaviour of a pure stellar component, suggesting 
the presence of hot ($\sim 300$K) dust. We tentatively argued that at least some K+A galaxies may be 
dusty starbursts (e.g. \citealt{Poggianti2000}), where dust is heated by young stars  in obscured super 
star clusters (responsible for the ultraviolet flux) and in the circumstellar envelopes of AGB stars (e.g.,
\citealt{Kelson2010,Chisari2012}).

In this paper we revisit some of these issues in the light of updated models by Maraston (2013, in preparation,
hereafter M2013; see \citealt{Noel2013} for a description) which assume less fuel for the AGB phase, {  and with the widely used models of \citealt{BC03} (henceforth BC03)}. Otherwise we use the same tools, photometry, and spectra as those used in Paper~I. We adopt the standard cosmological parameters of $\Omega_M=0.27$ and $\Omega_{\Lambda}=0.73$ with $H_0=71$km s$^{-1}$ Mpc$^{-1}$.

\section{Data and Models}

\begin{table}
\centering
\begin{center} 
\begin{minipage}{0.5\textwidth}
   \caption{\bf Ages for the {\tt STARLIGHT} and M2013 models}                                                   
   \begin{tabular}{ l  c  c }           
   \hline\hline                   
Bin Number   & {\tt STARLIGHT}  &  {\tt M2013}  \\
   \hline
    1& 1.00e+00& 1.00e+00 \\
    2& 3.16e+00& 3.00e+00 \\
    3& 5.01e+00& 5.00e+00  \\
    4& 6.61e+00& 6.50e+00 \\
    5& 8.71e+00& 8.50e+00 \\
    6& 1.00e+01& 1.00e+01 \\
    7& 1.45e+01& 1.50e+01 \\
    8& 2.51e+01& 2.50e+01 \\
    9& 4.00e+01& 4.00e+01 \\
  10& 5.50e+01& 5.50e+01 \\
  11& 1.02e+02& 1.00e+02 \\
  12& 1.61e+02& 2.00e+02 \\
  13& 2.86e+02& 3.00e+02 \\
  14& 5.09e+02& 5.00e+02  \\
  15& 9.05e+02& 9.00e+02 \\
  16& 1.28e+03& 1.00e+03 \\
  17& 1.43e+03& 1.50e+03 \\
  18& 2.50e+03& 2.00e+03 \\
  19& 4.25e+03& 4.00e+03 \\ 
  20& 6.25e+03& 6.00e+03 \\ 
  21& 7.50e+03& 7.00e+03 \\ 
  22& 1.00e+04& 1.00e+04 \\ 
  23& 1.30e+04& 1.30e+04 \\ 
  24& 1.50e+04& 1.50e+04 \\ 
%  25& 1.80e+04& 1.50e+04 \\
   \hline
   \end{tabular}
    \end{minipage}
   \label{ages}
   \end{center}
\end{table} 

The {\sc Starlight} code \citep{Cid2005} and its application to the K+A galaxies in our sample were fully described 
in Paper~I, to which we refer for further details. Briefly, we use the full star formation histories \footnote{ In Paper~I we binned the models into 3 broad age groups and 3 metallicity bins} 
provided by the {\sc Starlight} population synthesis models of the SDSS spectra. These star formation histories for 24 ages and 5  metallicities are combined to {\em predict} 
the spectral energy distributions (i.e., the {\it photometry}) from the GALEX FUV ($0.15\mu$m) to the WISE MIR ($22\mu$m) bands using the  M2013 models to generate synthetic SEDs, which are then compared with the observations.  

The dataset is detailed in Paper I, but the essential characteristics are: optical photometry (ugriz) and spectroscopy from
the SDSS \citep{Abazajian2009}; Vacuum UV data are from the GALEX imaging databases  \citep{Morrissey2007}; Infrared data
in (Y)JHK come from either the 2MASS  \citep{Skrutskie2006} or UKIDSS \citep{Lawrence2007}.
Mid-infrared data are taken from the Spitzer Heritage Archive or WISE \citep{Wright2010} datasets. A full description of the aperture matching criteria between these catalogs is given in Paper~I.

The observations are corrected for Galactic foreground extinction using the values of \cite{Schlafly2011} and for internal extinction within each galaxy using the values given by the {\sc Starlight} models. The SDSS input spectra to {\sc Starlight} were previously corrected for foreground extinction, so the internal extinction calculated in the population synthesis fits should, at least in principle, provide a reasonably reliable estimate of the average extinction affecting the overall stellar population within the fiber aperture. {  For both the internal and foreground extinction corrections we used a reddening curve typical of the Milky-Way as parametrized by \cite{Cardelli1989}, which here we refer to as either CCM when used in the {\sc Starlight} models, or MW when used to correct the photometric data.} 

{  We should point out that, as described in Paper~I, the {\sc Starlight} models use the BC03 \citep{BC03} stellar library to fit the SDSS spectra, so our synthetic SEDs built using the M2013 models (chosen to explore the putative signatures of AGB stars in the photometric data of our galaxies) could in principle be affected by systematic effects due to imperfect matching between the ages and metallicities of these two libraries. In order to quantify any systematic effects, therefore, we also computed synthetic SEDs using BC03 only}.  Our procedure introduces the relative fractions of stars in each age and metallicity bin in the Maraston and BC03 models by exploiting the extra information provided by the SDSS spectra, to {\it predict} the observable signatures of stellar populations, especially, in this case, the TP-AGB stars. 

It is important to emphasize that we do not {\em fit} the model SEDs to the observations: the only free parameter in our SED models is the wavelength-independent photometric normalization, for which we use the SDSS $i'$-band, and we examine how well these models reproduce the observations, in particular (but not solely) the expected IR contribution from TP-AGB stars. In Paper~I we found that our synthetic SEDs failed to predict the observed mid-IR fluxes, so we arbitrarily added a $T_{dust}=300^o$K Black-Body component to all  galaxies, which we normalized to the WISE $12\mu$m fluxes. 

%Table 1 presents  in the first column the 24 age bins calculated by {\sc Starlight}, and the corresponding age bins in the M2013 models (units are Myr).

The  M2013 {  and BC03 models} give populations for a wide range of ages and metallicities as summarized in
Tables 1 and 2. As shown in Table 1, it is possible to match reasonably well the  {\sc Starlight} ages
to the grid of M2013 without recourse to interpolation. The same is not the case for the metallicities,  for 
which the match was done as indicated in Table 2.  We verified that our synthetic SEDs are not sensitive to the choice of IMF by using both the Salpeter and Kroupa realizations of the M2013 models.
 
\begin{table*}
\centering
\begin{center}
\begin{minipage}{\textwidth}
  \caption{\bf Metallicity bins used to match {\tt STARLIGHT} and M2013 models }
  \begin{tabular}{ l l l }
  \hline\hline
 				& {\sc Starlight} 			& {\sc Maraston}								\\ 
Metallicity			& BC03+Chabrier 			& M2013+Kroupa		  					\\ \hline 
{\bf extremely poor}	& Z=(0.0001+0.0004)		&use logZ=-1.35 							\\
{\bf very poor}		& Z=0.004					&use logZ=-[1.35+0.58(logZ=-0.33)]/1.58 			\\
{\bf poor}			& Z=0.008					&use logZ=-[0.33+0.19(logZ=-1.35)]/1.19 			\\
{\bf solar}			& Z=0.02	 				&use logZ=0 								\\
{\bf rich}			& Z=0.05					&use logZ=+0.35%\tablenotemark{a}  	 			
\\ \hline  
   \end{tabular}
  % \tablenotetext{a}{Unfortunately the logZ=0.67 models are tabulated for a very different wavelength range.}
   \label{metals}
\end{minipage}
\end{center}
\end{table*}

The synthetic spectral energy distributions are then calculated as:

\begin{equation}
F(\lambda)= k {\sum_{i=1}^{24}}{\sum_{j=1}^5} S_{ij} M_{ij} (\lambda)
\end{equation}
\smallskip\\
{  where $M_{ij}$ are either the M2013 or BC03 models} for the $i=1,...,24$ age bins and $j=1,..,5$
metallicity bins, and $S_{ij}$ are the mass fractions returned by {\sc Starlight} for the relevant age and
metallicity bins (Tables 1 and 2).  The normalisation constant $k$ is independent of wavelength and is the only free parameter of the model.  

%Given the high metallicities of the intermediate-age components, in Paper~I we used a Milky-Way reddening law to correct both for foreground and internal extinction. This resulted in a significant excess over the models in the GALEX bands, which, in view of the presence of a hot dust component, we interpreted as evidence of hidden star formation despite the lack of strong  emission lines or 20cm flux \citep{Nielsen2012}. However, since the high star formation rates implied by the  mass fractions in the intermediate age population well qualify as a starburst, here we have relaxed the above constraint and assumed an extinction law more similar to that of local starbursts such as 30 Doradus (which is also similar to that of the Bar of the metal-poor Small Magellanic Cloud) using the parametrization of \cite{Misselt1999} that is specially well suited for the UV.

\section{Results}
\subsection{Comparison between the M2005 and the M2013 models}

\begin{figure*}
\includegraphics[height=12cm]{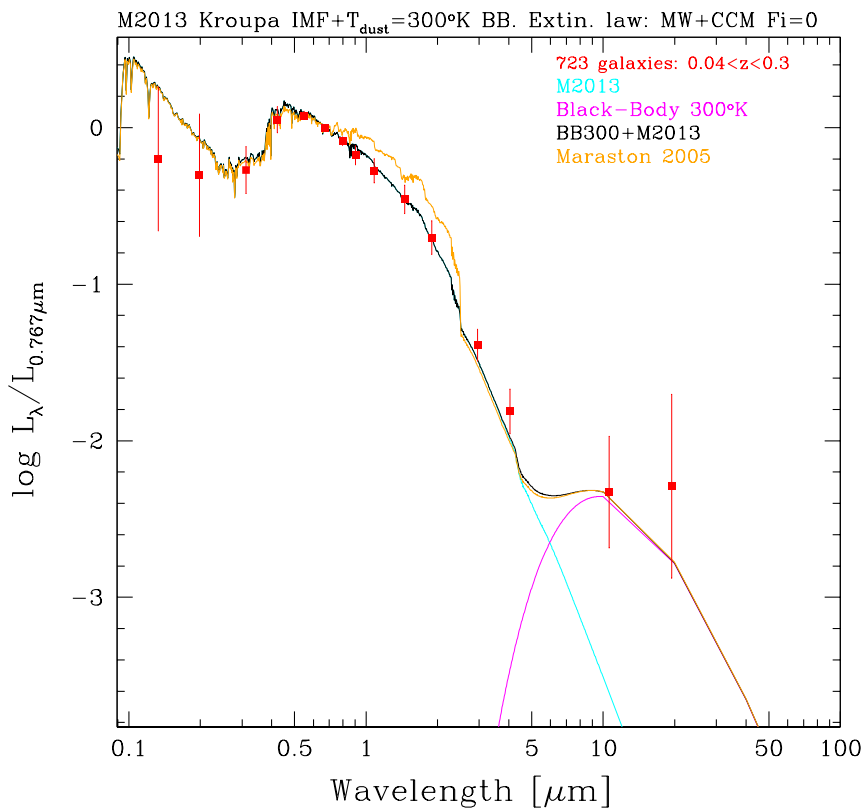} %\includegraphics[height=8cm]{propris18E0_BB300_2013.eps}
%\vspace{-2cm}

\caption {Comparison between the predicted synthetic SED for the average of the 808 K+A galaxies in our sample and the average of  the observed broad-band luminosities for a maximum of 723 galaxies (see Table~\ref{errors}) of redshifts $0.04<z<0.3$ (selected to minimise wavelength smearing in the averaged SEDs) corrected for internal and foreground extinction using the typical MW extinction law, and normalised to the SDSS $i'$-band ($F_{0.767\mu m}$) .  The orange line shows the synthetic SED constructed using the \cite{Maraston2005} models and the black line shows the result of using exactly the same stellar populations, but with the synthetic SED built using the \cite{Maraston2013} models.  The magenta line represents the $T_{dust}=300^oK$ Black-Body component that we added to match the MIR data.}
 \label{seds}
\end{figure*}

In order to remove the intrinsic variance in luminosity among the galaxies in the sample it is useful to normalise the fluxes to a fiducial wavelength, for which we chose the SDSS $i'$-band ($F_{0.767\mu m}$) with data for 800 galaxies. Figure~\ref{seds} presents the comparison between the predictions of the average synthetic SED computed using Equation~1 for the mean of the {\sc Starlight} stellar populations of all the galaxies in our sample, and the observations. The averaging of the observations was done for the rest-frame luminosities at each redshift and, in order to minimise the resulting smearing in the rest-frame wavelengths, only galaxies within the redshift range indicated in the figure legend (723 galaxies for $z=0.04-0.3$) were included in these averages. The error-bars show the $1\sigma$ dispersion of the distribution of normalised fluxes for each bin, not observational errors.
The orange line shows the SED predicted using the old Maraston (2005; M2005) models, and the black line shows the prediction of the new Maraston (2013; M2013) models.  Exactly the same {\sc Starlight} stellar populations were used to compute both SEDs.

 %Notice that the number of objects used in the averages is different for the different wavelengths and metallicity bins; the values quoted in the figure legends correspond to the number of objects detected in the WISE $12\mu$m band.  
%The average stellar populations for the sample are shown for three broad metallicity bins (chosen for clarity) in the insert of the panel.  The various colour-codes used in these figures are explained in the figure legends and in more detail in the figure caption.

%This figure illustrates the conclusion we reached in Paper~I that our models systematically over predict the observed UV fluxes. The figure also shows that in the UV galaxies with predominantly metal poor intermediate-age populations (red dots) are systematically more luminous as predicted by the models (Paper I). The same seems to be the case in the mid-IR bands, although here our models lack predictive power. % in these bands, which are dominated by our imposition of a Black-Body of constant temperature ($T_{dust}=300^o$K) for all galaxies. 
%The left panel shows the observed luminosities in Solar units corrected for extinction assuming a Milky-Way (MW) reddening law for both the foreground and internal extinction as in Paper~I. 

Clearly the AGB-heavy M2005 models significantly overestimate the contribution of TP-AGB stars to the near-IR luminosities of intermediate-age stellar populations, while the new `lighter' Maraston (2013; AGB-light) models with less fuel in the AGB phase adequately  reproduce the observed  near-IR fluxes of K+A galaxies. We recall that the M2013 models make no pretence of predicting the mid-IR fluxes, for which they simply extrapolate the expected blackbody fall-off of stellar atmospheres beyond the $K$-band as shown by the cyan line in Figure~\ref{seds}, and therefore that any discrepancy at wavelengths longer than $K$ is not relevant for these models. In Paper~I we found that the M2005 models systematically over-predict the UV luminosities of K+A galaxies. Figure~\ref{seds} shows that this feature remains (as expected) unchanged in the new models.  We remark that the average synthetic SED is constructed by first averaging the {\sc Starlight} stellar populations for the 808 galaxies in our sample, and then applying Equation~1 to combine the average stellar populations with the Maraston models.

\subsection{Comparison between M2013 and BC03 stellar libraries}

As shown in Figure~\ref{seds}, the new Maraston AGB-light models provide a much better fit than the M2005 models over a wide range of wavelengths, but not in the UV. It is interesting to compare the predictions of the M2013 models to those of `pure' BC03 synthetic SEDs. This is relevant not only for the contribution of AGB and post-AGB stars, but also to understand the discrepancy in the UV, which, as mentioned above, could be affected by our mixing of stellar libraries.

\begin{figure*}
\includegraphics[height=12cm]{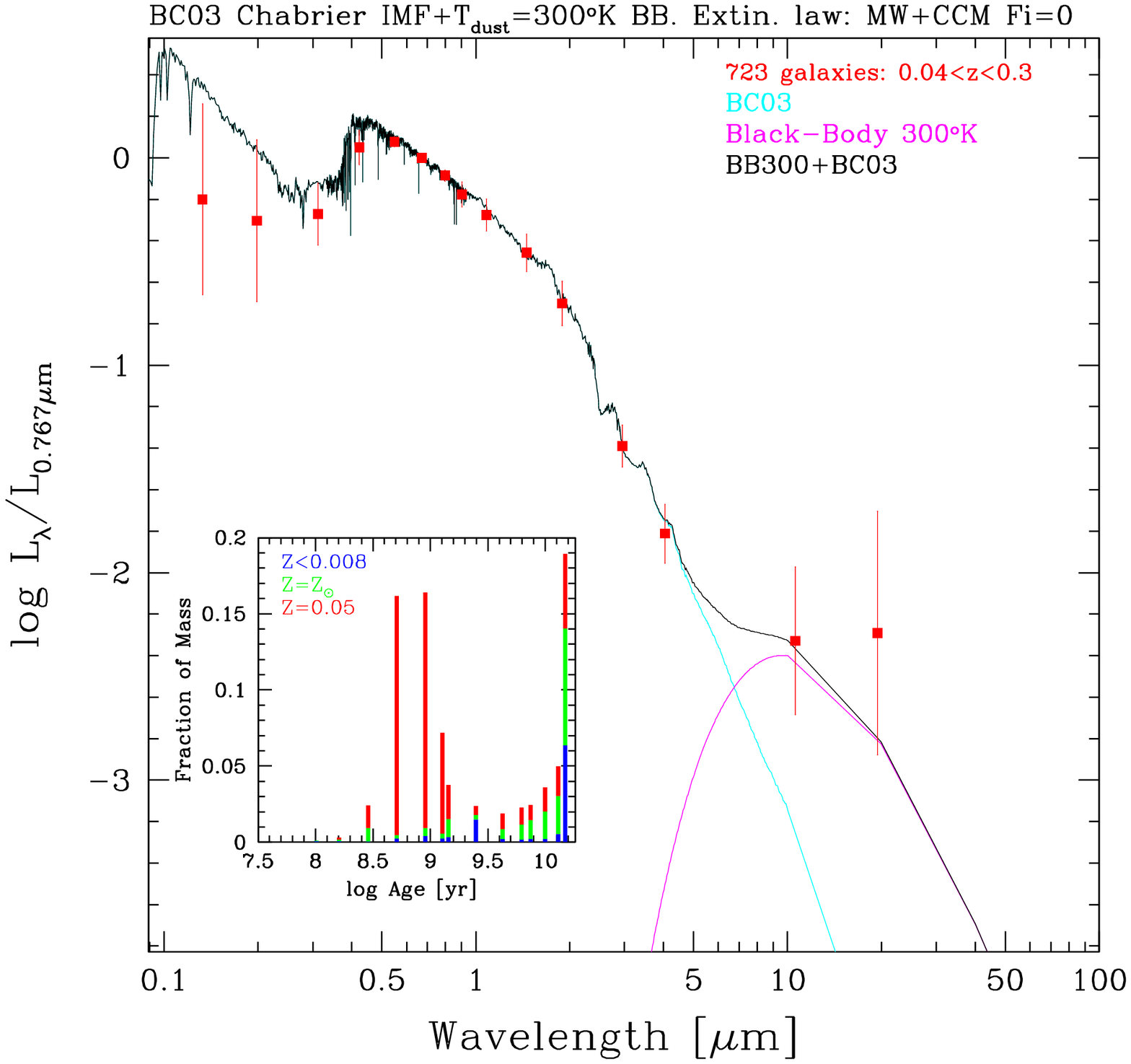} %includegraphics[height=8cm]{propris30E2_BB300.eps}
%\vspace{-2cm}
\caption {Comparison between the photometric observations and the mean synthetic SED built using the BC03 models only. The averaging of the photometric data points was done as for Figure~\ref{seds} as well as the corrections for extinction using the CCM reddening law.  {  The insert presents the average star-formation history - for three broad metallicity bins - of all galaxies in the sample used to compute the mean synthetic SEDs, for both the BC03 model shown in this figure, and the M2013 models shown in Figure~\ref{seds}. }}
 \label{norma}
\end{figure*}

Figure~\ref{norma} shows the synthetic SED built using the BC03 models only, for exactly the same age and metallicity bins as used in the {\sc Starlight} population synthesis fits. The data points are the same as those shown in Figure~\ref{seds}.  The figure shows that the synthetic SED built using the BC03 stellar library seems to reproduce equally well than M2013 the photometric observations from the optical to the near-IR. The BC03 models also provide predictions for the WISE $3.5\mu$ and $4.6\mu$ bands which, as we will see later, may provide useful additional diagnostics on the AGB populations. 

The UV discrepancy remains in the BC03 models. For both M2013 and BC03 we verified that this discrepancy is not due to our choice of reddening law: while the FUV fluxes increase significantly using a UV-strong reddening law, such as that of \cite{Gordon2003} for the SMC bar, the {\sc Starlight} populations also change with extinction law such as to make the synthetic SEDs steeper in the UV, thus cancelling the improvement resulting from the augmented extinction corrections in the UV photometry.

\subsection{Detailed object-by-object comparison between models and data}

\begin{figure*}
 \includegraphics[height=9cm]{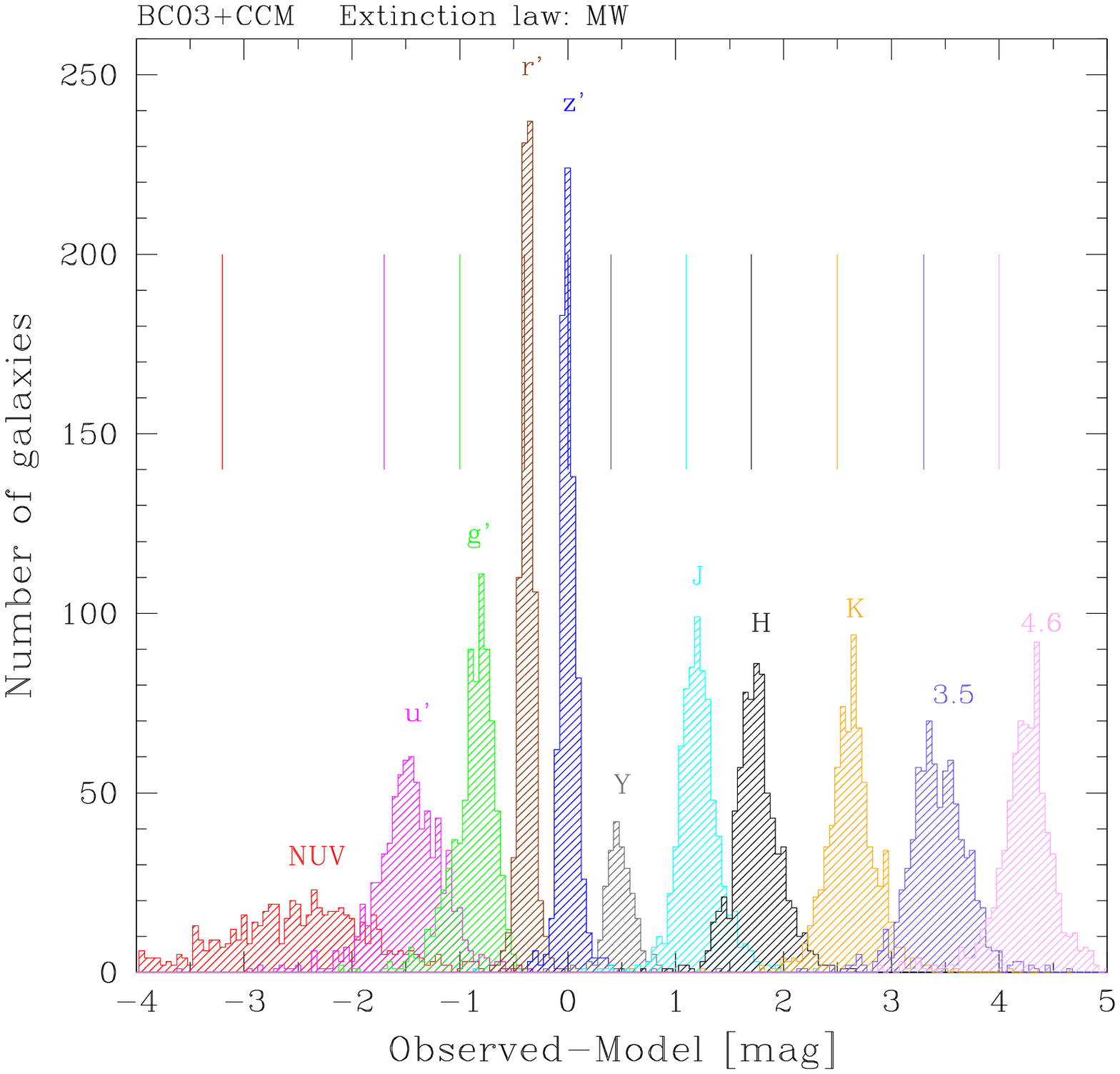}\includegraphics[height=9cm]{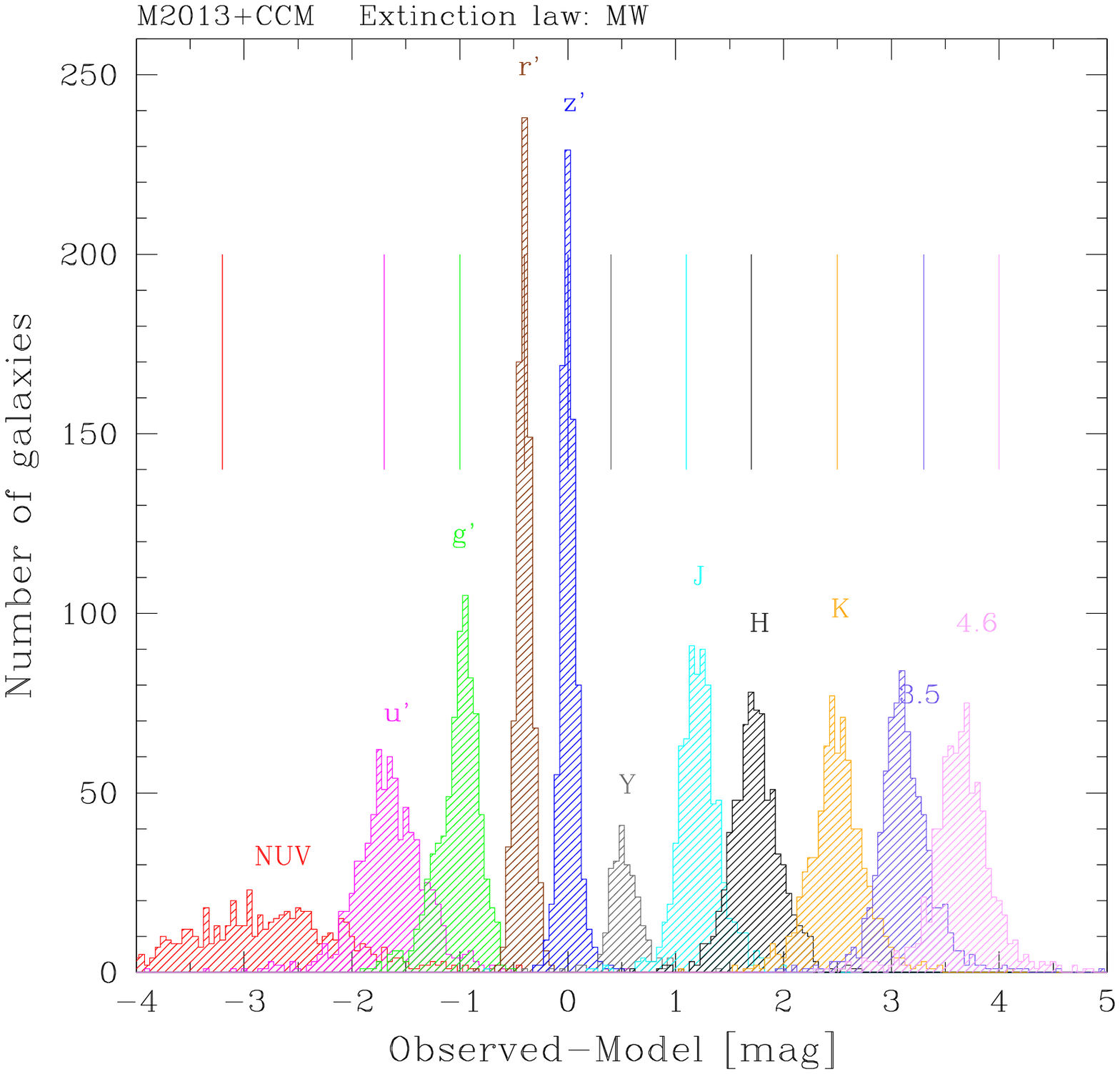}
 \caption{{\bf Left}. Distributions of the difference between the observed and modelled photometric fluxes from the FUV to the mid-IR for the BC033 models. (The details of each histogram are summarised in Table~\ref{errors}.) The histograms have been shifted for clarity and the vertical bars show the zero points for each color ($Observed-Model=0$).  {\bf Right.} Same as left but for the M2013 models.}
 \label{colores}
\end{figure*}

In order to better compare the predictions of our synthetic SED models, we show in Figure~\ref{colores} an object-by-object comparison between the models and the observations where for each band from the NUV to the mid-IR we plot the distribution of (Observed-Predicted) fluxes expressed in magnitudes {  for both sets of models}. For clarity we have omitted the FUV and the longest mid-IR bands in this figure, but the relevant parameters are listed in Table~\ref{errors} that summarises the fit parameters for the 15 bands and for both M2013 and BC03. 

%The figure and the table show that the agreement between the observed and predicted photometric fluxes from the optical to the near-IR is even better than in Figure~\ref{norma}, with the exception of the $Y$ and $J$ bands that show a mean systematic shift of about $0.1$ magnitudes. This shift is magnified on the right panel of Figure~\ref{colores}: the models predict systematically larger fluxes than observed.  %We will return to this in the following section.

Table~\ref{errors} shows several interesting results: M2013 seems to provide better fits in the FUV to $g'$ bands and also in the $K_S$-band, whereas both models fit equally well from $r'$ to $H$.  Both models over-predict that $Y$ and $J$ fluxes, while the BC03 models also systematically over-predict the $3.5\mu$ and $4.6\mu$ fluxes (for which M2013 lacks predictive power). % confirms the substantial disagreement between models and observations for the FUV and FUV bands. Interestingly, in these two bands the predictions of the average K+A stellar population (Figure~\ref{norma} right panel) are even better than in the object-by-object comparison, which is more affected by outliers. %The table also gives the dispersions in the observed fluxes normalised to the $i'$ band.

\begin{figure*}
\includegraphics[height=7.5cm]{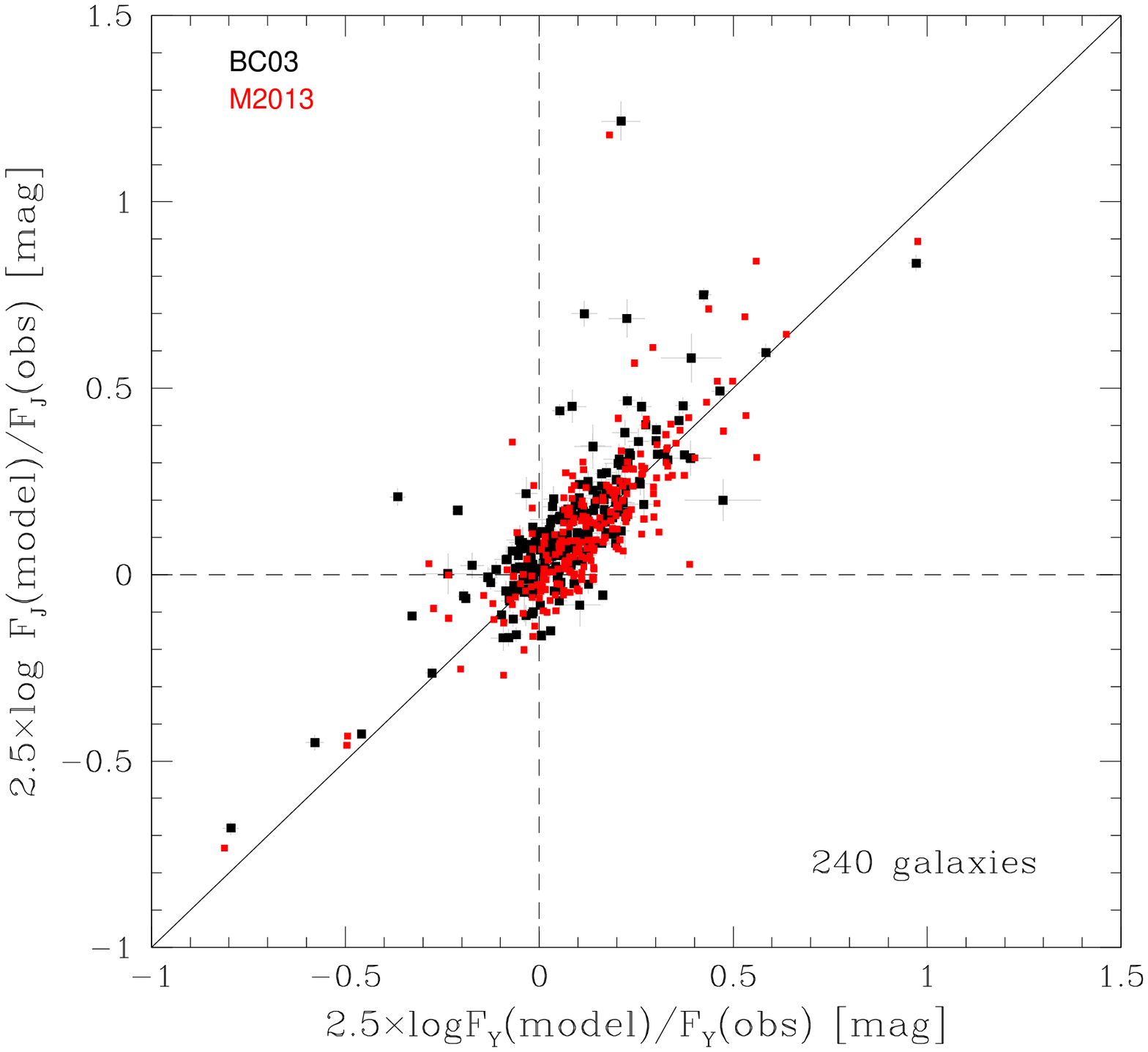}\includegraphics[height=7.5cm]{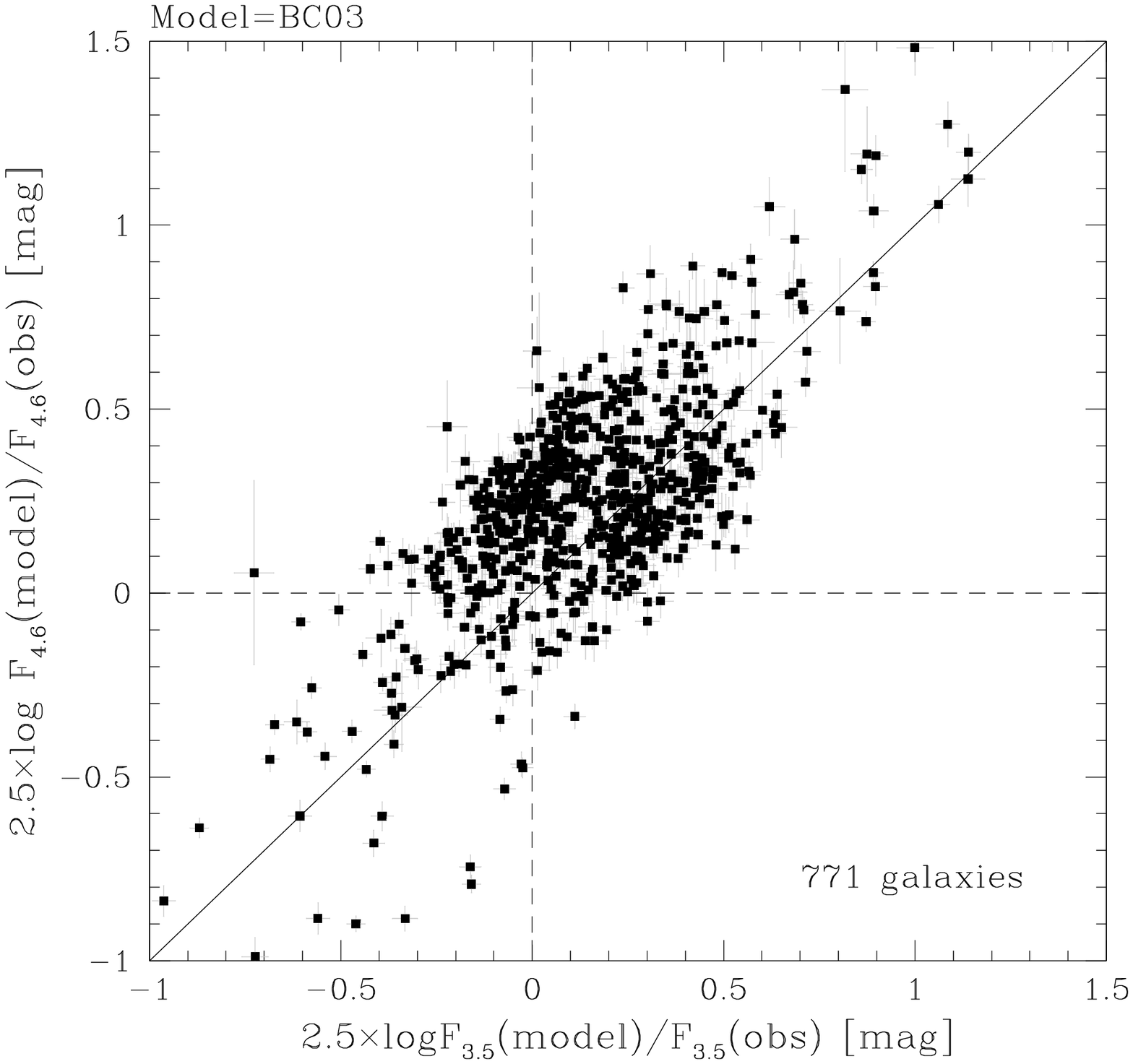}
 \caption{ {\bf Left}. Object-by-object comparison between models and observations for the UKIDSS $Y$ and $J$ bands for the BC03 models (black points) and the M2013 models (red points). {\bf Right}. Object-by-object comparison between models and observations for the WISE $3.5\mu$ and $4.6\mu$ bands. The dashed lines mark the zeroes (Model=Observed), and the solid line shows the locus of Y=X. }
\label{cacao}
\end{figure*}

\begin{table*}
\centering
\begin{center}
\begin{minipage}{\textwidth}
  \caption{\bf Detailed Comparison between Models and Observations}
  \begin{tabular}{l c c c c c c } 
  \hline\hline\small
                         &  \multicolumn{4}{c} {(Observed-Model) [mag]}  & \multicolumn{2}{c} { } \\   
 			 & \multicolumn{2}{c}{\bf BC03} & \multicolumn{2}{c}{\bf Maraston 2013}    & Number of  & Mean obs.   \\ 
Band			&  Average 	& Dispersion  & Average 	 & Dispersion 	               &  galaxies     & error  [mag]	\\ \hline
FUV	&  1.55	&  1.51	&  1.27	&  1.47	&  209	&  0.32 \\ 
NUV	&  0.67	&  0.81	&  0.35	&  0.86	&  609	&  0.27 \\ 
 u'	&  0.24	&  0.35	&  0.05	&  0.36	&  766	&  0.13 \\ 
 g'	&  0.12	&  0.22	& -0.02	&  0.21	&  774	&  0.06 \\ 
 r'	&  0.02	&  0.08	&  0.00	&  0.08	&  775	&  0.05 \\ 
 i'	&  0		&  0		&  0		&  0		&  777	&  0.07 \\ 
 z'	&  0.00	&  0.09	&  0.01	&  0.09	&  772	&  0.06 \\ 
  Y	&  0.08	&  0.17	&  0.13	&  0.18	&  240	&  0.02 \\ 
  J	&  0.11	&  0.22	&  0.12	&  0.22	&  755	&  0.10 \\ 
  H	&  0.05	&  0.23	&  0.04	&  0.24	&  749	&  0.13 \\ 
  K	&  0.13	&  0.24	&  0.00	&  0.27	&  740	&  0.13 \\ 
3.5	&  0.14	&  0.29	& -0.18	&  0.28	&  768	&  0.03 \\ 
4.6	&  0.25	&  0.30	& -0.35	&  0.31	&  771	&  0.05 \\ 
 12	&  --- 	&  --- 	&  ---	        &  ---	        &  611	&  0.24 \\ 
 24	&  ---	        &  ---  	& ---		&  ---		&  225	&  0.13 \\ 
			
\hline  
   \end{tabular}
    \label{errors}
\end{minipage}
\end{center}
\end{table*}
{  The mean errors of the average values listed in Table~\ref{errors} would be the dispersions divided by $\sqrt{N-1}$. Thus}, because of the large number of galaxies observed in each band, these systematic discrepancies are highly significant as illustrated in Figure~\ref{cacao}, which compares model and observed fluxes for four of the discrepant bands. {  The right panel on this figure shows that}, with some differences for individual objects, these discrepancies are present in both the M2013 and BC03 models. %Table~\ref{errors} show that these discrepancies are present in both M2013 and BC03 models

We should also notice the discrepancy between the predictions of the BC03 and M2013 models for the SDSS $u'$ and $g'$ bands: while M2013 agrees with the observations within the errors, BC03 significantly over-predict the observed fluxes. For consistency with Paper~I, the comparisons of Table~\ref{errors} were done using monochromatic fluxes at the effective wavelengths of the corresponding filters, most of which are relatively free of strong absorption features except for $g'$, where the 4000\AA\ break is at the edge of the band and contributes to the integrated fluxes even for our lowest redshift galaxies (for $z\leq0.3$ the Lyman-break remains outside of the FUV band). 

We calculated the `k-corrections'  for the 15 bands by integrating the redshifted model fluxes over the 15 filter passbands. As expected, the effect is significant ($>0.04$ mag) only for the $g'$ band, for which the mean (Observed-Model) for BC03 is reduced by 0.08 mag with respect to the monochromatic values of Table~\ref{errors}, 
%For $g'$ the mean is reduced to 0.06 mag while for FUV it is increased to 1.48 mag: the continuum slope more than compensates the effect of  $Ly\alpha$. 
while the dispersions in all bands remain essentially unchanged.  We will return to the UV discrepancy in the next section.

%\end{document}

\section{Discussion}
\subsection{Synthetic SEDs from SDSS spectra}

In Paper~I we asked the question: can we use the stellar populations derived from SDSS spectra to predict the photometric SEDs of post-starburst galaxies from the FUV to the mid-IR?  Our study was based on publicly available models and observations; we used the library of population synthesis models for about 700,000
SDSS galaxies compiled by \cite{Cid2005}, which is publicly available at {\tt www.starlight.ufsc.br}.  Because we were mainly interested in exploring the effects of intermediate-age TP-AGB stars in the integrated IR colours of galaxies, we chose to use the \cite{Maraston2005} stellar library for the reasons explained in the Introduction. %that was specifically built to deal with these types of stars. 

The answer to our original question was clear: No. Our synthetic SEDs predicted too much UV light, too much near-IR flux, and far too little flux for the 
WISE $12\mu$ and $24\mu$ mid-IR bands. While the latter could be accounted reasonably well by adding an ad-hoc hot dust component, 
we found no simple explanation for the other discrepancies. In this paper we investigated the same question using the new AGB-light models of Maraston, and also the BC03 models used for the {\sc Starlight} fits, finding that with these libraries we can indeed build synthetic SEDs that reproduce remarkably well the observed broad-band photometric colours of K+A galaxies from the optical to the mid-IR {\em without free parameters}.  

Independently of the stellar library, however, our new models still predict too much flux in the UV compared to the GALEX observations. We verified that the systematic discrepancy is not due to a particular choice of reddening law, which implies that it would be rather unlikely that the large scatter in the observed UV properties of K+A galaxies be due to differential reddening effects. 

As mentioned above, the excellent agreement between the synthetic SEDs and the photometric observations in the optical and IR bands endows the models with a predictive power such that even small discrepancies with the observations need to be understood. Table~\ref{errors} documents these discrepancies. For the M2013 library we have significant discrepancies in the FUV, NUV, $u'$, $Y$, and $J$ bands, whereas, in addition to these bands, for the BC03 models  we find discrepancies in the $K_S$, $3.5\mu$ ($L$) and $4.6\mu$ ($M$) bands. In the following sections we explore the possible origin of these discrepancies.

%Thus, while at optical and the short mid-IR wavelengths K+A galaxies are a remarkably homogenous class of objects, this does not seem to be not the case in the UV (as  is also not the case in the longer mid-IR bands although our simplistic models make no pretence of reproducing the observations there). Also, the scatter in the {\em observed} fluxes at $12\mu$ and $24\mu$ is actually comparable or even larger than in the UV.  

%Our new answer to the original question therefore is: Yes in the optical and near-IR; No in the UV and mid-IR. However, the fact that {\em without free parameters} our models reproduce the observed broad-band colours of K+A galaxies to substantially better than $0.1$ mag (except in $Y$ and $J$) allows us to use them as a starting point to explore the possible causes for the discrepancies we observe in the UV and the mid-IR.

\subsection{Intermediate-age AGB stars}

The massive AGB stars that should be present in K+A galaxies (from the stellar evolution of their younger stellar
populations) should contribute to dust heating and provide an explanation for the apparent excess at mid-infrared
wavelengths over the Rayleigh-Jeans tail of the stellar blackbodies seen in both sets of models. Massive AGB stars are observed as either 
Carbon (C) stars or Oxygen (M) stars depending on the surface C/O abundance ratio during the thermal-pulsing
phase \citep{Renzini1981}. 

Population studies in nearby galaxies \citep{Mouhcine2003,Groenewegen2007} show 
that  $N_C/N_M$,  the ratio of the number of C stars to the number of M stars, is a strong function of the 
parent galaxy metallicity [$Fe/H$]. Metal-rich galaxies (like M31 or the Milky-Way) contain predominantly M stars, 
whereas the AGB populations of metal poor galaxies, like the SMC and other dwarf irregulars in the Local Group, are
dominated by Carbon stars. This would lead us to expect that the populations of AGB stars in the intermediate-age
component of K+A galaxies, which for the most part have solar or greater metallicities, should be completely dominated by 
late-type M stars.

In fact this assumption is consistent with our observations. For example, the circumstellar envelopes (CSE) of 
Oxygen-rich AGB stars (O-CSE) have spectral energy distributions in the mid-IR that are remarkably 
similar to those we observe in K+A galaxies, particularly, but not exclusively, for stars with optically-thick 
envelopes \citep{Bedijn1987,Sargent2010}. If the O-CSE are optically-thick, their contribution to the 
optical-infrared colours and spectra would be reduced (in agreement with claims by \citealt{Kriek2010} and
\citealt{Zibetti2013}). Indeed, \cite{Jimenez2005} find that about 1/3 of Galactic OH/IR stars lack an optical and 
even an infrared counterpart. But also for optically-thin envelopes the near-IR features of M stars are significantly 
less prominent than those of C stars \citep{Wright2009}. This may account for their weakness in the spectra by
\cite{Zibetti2013}. Moreover, the mid-IR SEDs of C-CSE are bluer (hotter) than those of O-CSE consistent with 
our observations. 

The spectra of late-type M stars exhibit deep VO and TiO molecular absorption bands in
the $1.05 - 1.15$ $\mu$m wavelength range, precisely in the $Y$ and $J$ bands where we find a significant
discrepancy between our models and the observations of K+A galaxies. Unfortunately, the spectra of
\cite{Zibetti2013} do not cover these wavelengths, but if indeed K+A galaxies contain dominant 
O-rich AGB populations we should expect these absorption bands to show-up prominently in their spectra.  

The BC03 models over-predict the $K_S$-band fluxes, which is not the case for the M2013 models that would seem, therefore, to provide a better treatment for the AGB phases. While M giants have prominent CO features in the $K_S$-band, even for the nearest galaxies in our sample the CO band-head at $2.3\mu$ is redshifted out of the 
$K_S$ filters of both UKIDSS and 2MASS, so it is unlikely that the discrepancy be due to CO absorption bands.

The BC03 models also predict too much flux in the $L$ and $M$ bands. M-giants have deep CO lines in the $M$ band, but not at $3.5\mu$ \citep{Plez2005}. There is evidence from Galactic M-giants and supergiants that the CO lines seem to become weaker with increasing metallicity \citep{Comeron2004}, which would favour Carbon stars as a possible explanation of the discrepancy at $L$ and $M$. Unfortunately, observations of metal-rich cool giants in these bands are rather scarce, which makes it difficult to test this hypothesis empirically.

%These stars also have strong CO lines in the $K_S$ and $M$ bands, which may explain why our models over predict the fluxes in these bands.  

\subsection{The UV Discrepancy}

%We have no simple explanation for the large variation in the normalised UV and mid-IR fluxes that we observe within K+A galaxies as a class.

\begin{figure*}
\includegraphics[height=10cm]{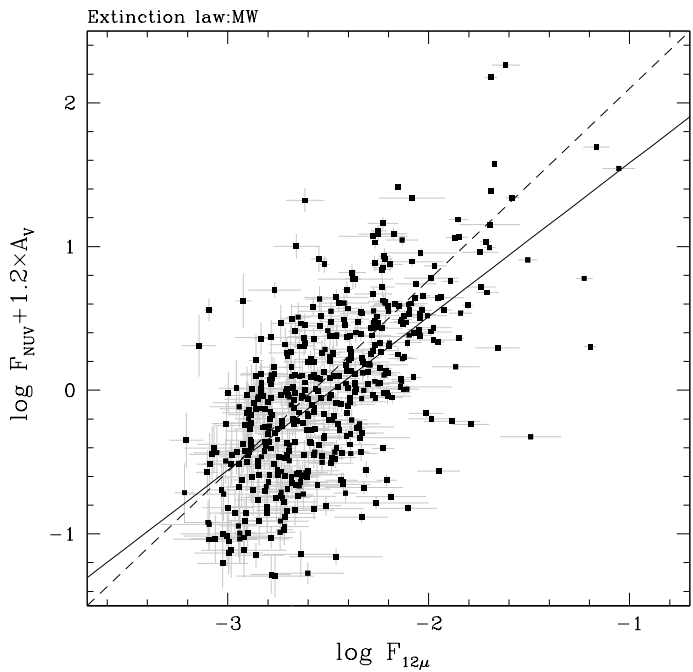}
\caption{ Plot of the observed NUV flux corrected for extinction, and normalised to the average $i'$ band flux of the sample, versus the observed $12\mu$ flux with the same normalisation. The solid line shows a least-squares fit to all the points, and the dashed line corresponds to X=Y.}
\label{exes}
\end{figure*}

Despite rather large differences in luminosity, the SEDs of K+A galaxies from the optical to the mid-IR below $5\mu$ are remarkably similar. This suggests that the large scatter we observe in the UV and MIR colours of these galaxies (Table~\ref{errors}) may have a common origin.  Indeed, Figure~\ref{exes} shows that NUV and $12\mu$ fluxes are correlated.  Since our models systematically {\em over predict} the UV fluxes, this correlation suggests that both the UV and mid-IR emission from these galaxies may arise from the same (short-lived) stellar populations. In this scenario stochastic effects would explain both the variance in UV and mid-IR fluxes, as well as the systematic discrepancy between models and observations in the UV.  

This happens because simple stellar populations (ssp) models, such as M2013 and BC03, integrate the stellar libraries over an IMF (in our case Kroupa or Chabrier) and therefore systematically over-sample the progenitors of rare stellar populations by including fractions of stars in the integrations. In particular, even for the relatively massive intermediate-age populations in K+A galaxies, the number of hot post-AGB stars present at any given time will be small and their observations, therefore, subject to significant stochastic effects.
%As we suspected but not expected, there is a strong correlation between the {\em observed} fluxes in the NUV and $12\mu$ bands.  This correlation is not {\it a priori} expected if the mid-IR emission arrises from hot circumstellar dust envelopes of (metal rich) AGB stars. In this scenario, the extra UV emission would have to come from hot post-AGB stars but our problem is that we already predict too much UV flux even without the contribution of hot intermediate-age post-AGB stars.  

An alternative explanation could be differential extinction: if K+A galaxies evolve from anything like the starburst galaxies observed at higher redshifts, in the post-starburst phase they could retain some of this clumpiness. The extinction seen by {\sc Starlight} in the SDSS spectra is some sort of ensemble average over the integrated stellar populations leaving, at least in principle, open the possibility for substantial spatial fluctuations in extinction. Clumpiness, however, introduces an extra degree of freedom in the modelling and one over which we have very limited direct observational constraints, if any. Therefore, while more in-depth modelling is clearly needed to elucidate the UV and mid-IR properties of post-starburst galaxies, we also need spatially resolved observations of a significant sample of objects if we want to constrain both spatial and temporal stochastic effects.  

\section{Conclusions}
%The following conclusions are robust and should not depend on our choice of stellar libraries or techniques to model the stellar populations in K+A galaxies:
Post-starburst galaxies provide ideal laboratories to investigate the properties of evolved intermediate age stellar populations at super-solar metallicities, which are extremely rare in our own Galaxy. In particular, these galaxies are ideal to search for the effects of TP-AGB stars in the integrated colours and spectra of galaxies at redshifts of $z=2-3$, when the age of the Universe was comparable to the typical ages of TP-AGB stars. Our findings are summarised as follows:

\begin{enumerate}

\item K+A galaxies are a remarkably homogeneous class of objects when viewed at optical and IR wavelengths, but not in the ultraviolet or the mid-IR beyond $5\mu$;

\item The synthetic SEDs derived from population synthesis fits to the SDSS spectra reproduce remarkably well the observed broad-band SEDs of K+A galaxies, in particular in the SDSS bands. This means, in particular, that the SDSS $3''$ spectral fibers provide a fair sampling of the overall stellar populations of the K+A galaxies in our sample;

\item The synthetic SEDs built using either the M2013 or the BC03 stellar libraries reproduce the broad-band IR colours very well, although small but significant discrepancies remain in the $Y$ and $J$ bands for both libraries and additionally for the $K_S$, $L$, and $M$ bands for BC03. We suggest that these discrepancies are likely due to insufficient modelling of metal-rich AGB and post-AGB stars in these libraries;

\item Population synthesis modelling of SDSS galaxies provides a robust method for building synthetic templates for measuring photometric redshifts of moderate redshift galaxies. However, it is essential to check that the templates also work in the UV when applied to galaxies at $z>1.5$.

\end{enumerate}

These conclusions can be directly tested observationally through IR spectroscopy of metal-rich K+A galaxies, which should reveal prominent VO 
and TiO molecular features in the $Y$ and $J$ bands, strong CO lines in the $M$, and possibly also $L$, bands, and a strong Silicate feature at 10$\mu$m.

Our models fail to reproduce the large variance in the integrated UV and mid-IR colours of K+A galaxies and systematically over predict the UV fluxes. Significant progress in understanding these issues will require spatially resolved observations of a reasonable sample of galaxies from the UV to the mid-IR, as well as a better understanding of AGB and post-AGB giants at high metallicity. In this context it seems particularly important to develop population-synthesis models based on Monte Carlo techniques explicitly taking into account stochastic effects introduced by short-lived stellar populations, for which ssp models are clearly inadequate. 

\section*{Acknowledgements}

We are grateful to Claudia Maraston for providing us with the M2013 models ahead of publication, and for several very useful comments to 
the manuscript version of this paper, and to Enzo Brocato for pointing out the importance of stochastic effects starting at the IMF. We thank our anonymous referee for many suggestions leading to a much improved version of the paper.

%\newpage

%\bsp
%\label{lastpage}
 
\end{document}